\documentclass[10pt]{article}
\usepackage{graphicx}
\usepackage{amsmath}

\def\Title#1{\begin{center} {\Large #1 } \end{center}}
\def\Author#1{\begin{center}{ \sc #1} \end{center}}
\def\Address#1{\begin{center}{ \it #1} \end{center}}

\newcommand\pubblock{\rightline{\begin{tabular}{l} Proceedings of the Second Annual LHCP\\ \pubnumber\\
         \pubdate  \end{tabular}}}

\newenvironment{Abstract}{\begin{quotation} \begin{center} 
             \large ABSTRACT \end{center}\bigskip 
      \begin{large}}{\end{large} \end{quotation}}

\newenvironment{Presented}{\begin{quotation} \begin{center} 
             PRESENTED AT\end{center}\bigskip 
      \begin{center}\begin{large}}{\end{large}\end{center} \end{quotation}}

\def\beq{\begin{equation}}
\def\eeq#1{\label{#1}\end{equation}}
\def\eeqn{\end{equation}}

\def\beqa{\begin{eqnarray}}
\def\eeqa#1{\label{#1}\end{eqnarray}}
\def\eeqan{\end{eqnarray}}

\let\bar=\overbar

\def\Dslash{\not{\hbox{\kern-4pt $D$}}}
\def\dslash{\not{\hbox{\kern-2pt $\del$}}}

\def\msb{{\bar{\ssstyle M \kern -1pt S}}}

\newcommand{\microbarn}{\mathrm{\,\mu b}}
\newcommand{\nanobarn}{\mathrm{\,nb}}
\newcommand{\invnb}{\mathrm{\,nb}^{-1}}

\newcommand{\tev}{\mathrm{\,TeV}}

\newcommand{\gevc}{{\mathrm{\,GeV\!/}c}}

\newcommand{\pt}{p_{\mathrm{T}}}
\newcommand{\BF}{\mathcal{B}}
\newcommand{\jpsi}{J/\psi}
\newcommand{\YoneS}{\Upsilon(1S)}
\newcommand{\plead}{p\mathrm{Pb}}
\newcommand{\pA}{p\mathrm{A}}

\usepackage{color}

\newcommand\pubnumber{ }

\newcommand\pubdate{\today}

\def\affiliation{
On behalf of the LHCb collaboration, \\
Key Laboratory of Particle \& Radiation Imaging (Tsinghua University), \\
Ministry of Education; \\
Center for High Energy Physics, Department of Engineering Physics, \\
Tsinghua University, Beijing 100084, China }

\begin{document}

\large
\begin{titlepage}
\pubblock

\vfill
\Title{  Quarkonia production in proton-lead collisions at LHCb  }
\vfill

\Author{ Yiming Li  }
\Address{\affiliation}
\vfill
\begin{Abstract}

The production of $\jpsi$ and $\YoneS$ mesons decaying into dimuon final state is studied at the LHCb experiment, with rapidity 1.5 $<$ y $<$ 4.0 or $-$5.0 $<$ y $< -2.5$ in proton-lead collisions at a nucleon-nucleon centre-of-mass energy $\sqrt{s_{NN}} = 5 \tev $, based on a data sample corresponding to an integrated luminosity of 1.6 $\invnb$. The nuclear modification factor and forward-backward production ratio are determined for prompt $\jpsi$, $\jpsi$ from $b$-hadron decay and $\YoneS$ mesons in study of the cold nuclear matter effects.

\end{Abstract}
\vfill

\begin{Presented}
The Second Annual Conference\\
 on Large Hadron Collider Physics \\
Columbia University, New York, U.S.A \\ 
June 2-7, 2014
\end{Presented}
\vfill
\end{titlepage}
\def\thefootnote{\fnsymbol{footnote}}
\setcounter{footnote}{0}
%

\normalsize 


\section{Introduction}
Heavy quarkonia states are good probes to quark-gluon plasma (QGP) 
in ultra-relativistic heavy-ion collisions, since their production is 
suppressed with respect to $pp$ collisions due to the formation of QGP. 
This effect is difficult to disentangle
from cold nuclear matter (CNM) effects, such as nuclear shadowing and energy loss 
of heavy quark (pairs), which also cause suppression of quarkonia production.
The proton-nucleus ($p$A) collisions provide unique environment to study CNM 
effects with the absence of QGP.

The LHCb detector collected a data sample corresponding to an
integrated luminosity of $1.6\,\invnb$ in $\plead$ collisions at the LHC.
The centre-of-mass energy of the $NN$ system $\sqrt{s_{NN}}$ 
is around $5\,\tev$.
Two beam configurations are used to cover different rapidity regions: 
the forward and the backward. The forward region corresponds to $1.5 < y < 4.0$ 
in the $NN$ center-of-mass frame, with the proton beam pointing towards the detector;
on the contrary the proton beam pointing away from the detector corresponds to
the backward rapidity region, with $-5.0 < y < -2.5$ in the $NN$ rest frame.
Using the data sample LHCb studies the production of $\jpsi$\cite{Aaij:2013zxa} 
and $\Upsilon$ mesons\cite{Aaij:2014mza} in $\plead$ collisions at $\sqrt{s_{NN}} = 5\,\tev$.

\section{Production cross-section of $J/\psi$ and $\Upsilon$ mesons} 
Both $\jpsi$ and $\Upsilon$ mesons are reconstructed with dimuon final states.
The trigger and selection efficiencies are determined with data-driven methods
wherever possible. 

Thanks to the excellent vertexing performance, 
the LHCb detector is able to distinguish prompt $\jpsi$ mesons and 
those from $b$ decays according to their pseudo-proper time distribution.
The integrated production cross-sections with $\pt < 14 \gevc$
are determined to be:
\begin{align*}
\sigma(\mathrm{prompt}\, J/\psi, 1.5 < y < 4.0) &= 1168 \pm 15 \pm 54 \microbarn, \\
\sigma(\mathrm{prompt}\, J/\psi, -5.0 < y < -2.5) &= 1293 \pm 42 \pm 75 \microbarn, \\
\sigma(J/\psi\, \mathrm{from}\, b, 1.5 < y < 4.0) &= 166.0 \pm 4.1 \pm 8.2 \microbarn, \\
\sigma(J/\psi\, \mathrm{from}\, b, -5.0 < y < -2.5) &= 118.2 \pm 6.8 \pm 11.7 \microbarn,
\end{align*}
where the first uncertainty is statistical and the second is systematic, 
as followed in the rest of the proceeding. The differential production 
cross-sections of prompt $\jpsi$ and $\jpsi$ from $b$ decay as functions
of $\pt$ or $y$ are also measured.

For $\Upsilon$ mesons with $\pt < 15 \gevc$, the integrated production 
cross-section times the decay branching fraction are measured with limited statistics:
\begin{align*}
\sigma(\Upsilon(1S), -5.0 < y < -2.5) \times \BF(1S) &= 295 \pm 56 \pm 29 \nanobarn,\\
\sigma(\Upsilon(2S), -5.0 < y < -2.5) \times \BF(2S) &= 81 \pm 39 \pm 18 \nanobarn,\\
\sigma(\Upsilon(3S), -5.0 < y < -2.5) \times \BF(3S) &= 5 \pm 26 \pm 5 \nanobarn,\\
\sigma(\Upsilon(1S), 1.5 < y < 4.0) \times \BF(1S) &= 380 \pm 35 \pm 21 \nanobarn,\\
\sigma(\Upsilon(2S), 1.5 < y < 4.0) \times \BF(2S) &= 75 \pm 19 \pm 5 \nanobarn,\\
\sigma(\Upsilon(3S), 1.5 < y < 4.0) \times \BF(3S) &= 27 \pm 16 \pm 4 \nanobarn,
\end{align*}
where $\BF(nS)$ denotes the branching fraction of $\Upsilon(nS) \to \mu^+\mu^-$ 
$(n = 1, 2, 3)$ decay. All the measurements above assume that $\jpsi$
and $\Upsilon$ mesons are produced with no polarisation.

\section{Cold nuclear matter effects on $J/\psi$ and $\Upsilon$ production}
The cold nuclear matter effects can be charaterised by 
nuclear modification factor
$R_{\pA} \equiv \frac{1}{A}\cdot(\frac{d\sigma_{\pA}}{d y})/(\frac{d\sigma_{pp}}{d y})$ and the forward-backward production ratio
$R_{\mathrm{FB}} \equiv \sigma_{\pA}(+|y|)/\sigma_{\pA}(-|y|)$.
With the production cross-sections measured in forward and backward rapidity
regions as introduced in the previous section, the CNM effects on 
prompt $\jpsi$, $\jpsi$ from $b$ decay and $\YoneS$ mesons 
can be studied ($\Upsilon(2S)$ and $\Upsilon(3S)$ yields are too low).
To determine the nuclear modification factor it requires 
cross-sections at $\sqrt{s} = 5 \tev$ in $pp$ collisions as a reference,
which is obtained by interpolating from $\jpsi$ and $\YoneS$ 
productions at 2.76 $\tev$, 7 $\tev$ and 8 $\tev$ $pp$ collisions 
measured by LHCb~\cite{Aaij:2012asz,Aaij:2014nwa,Aaij:2011jh,LHCb:2012aa,Aaij:2013yaa}.

\begin{figure}[htb]
\centering
\includegraphics[width=0.4\textwidth]{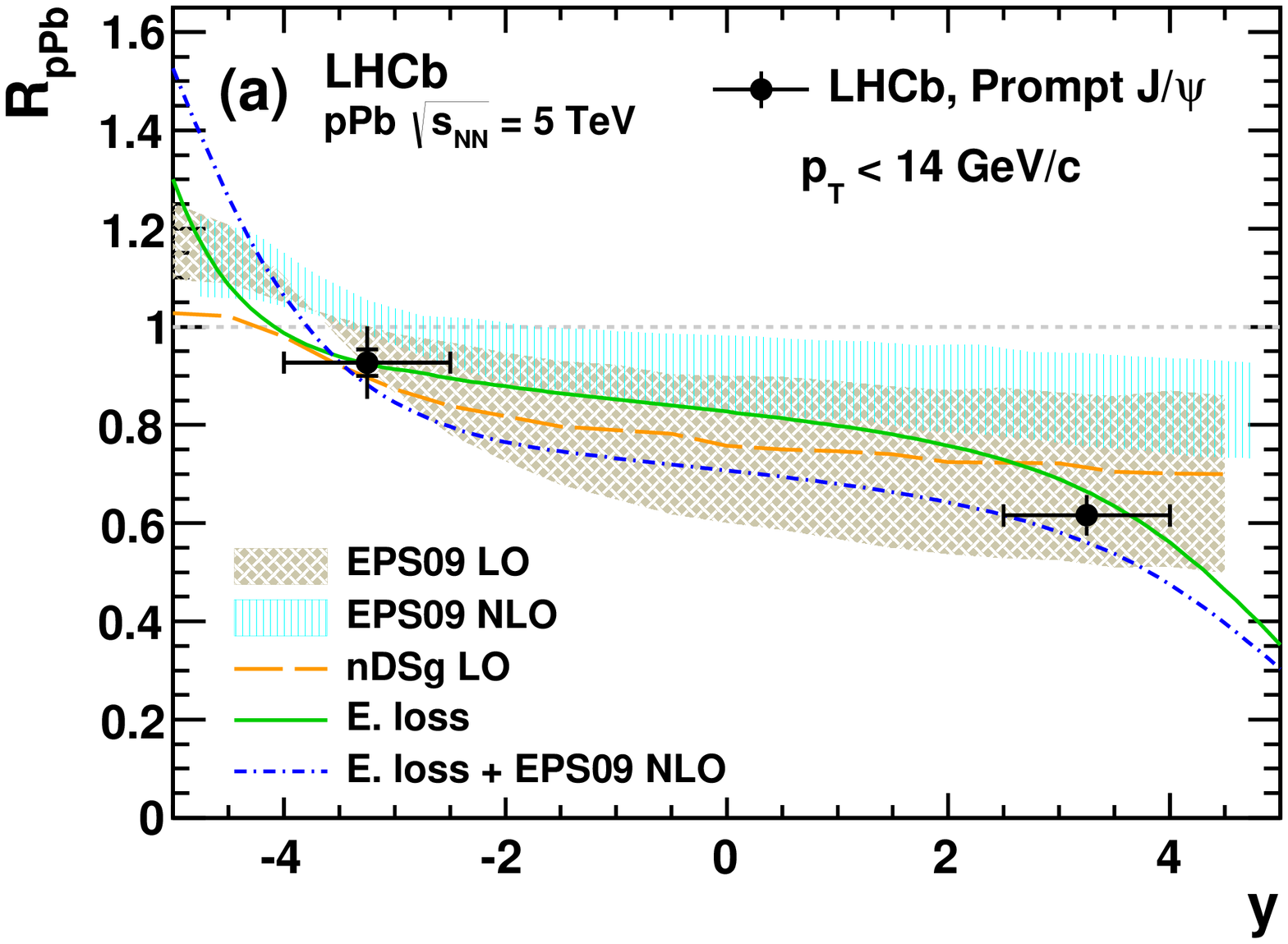}
\includegraphics[width=0.4\textwidth]{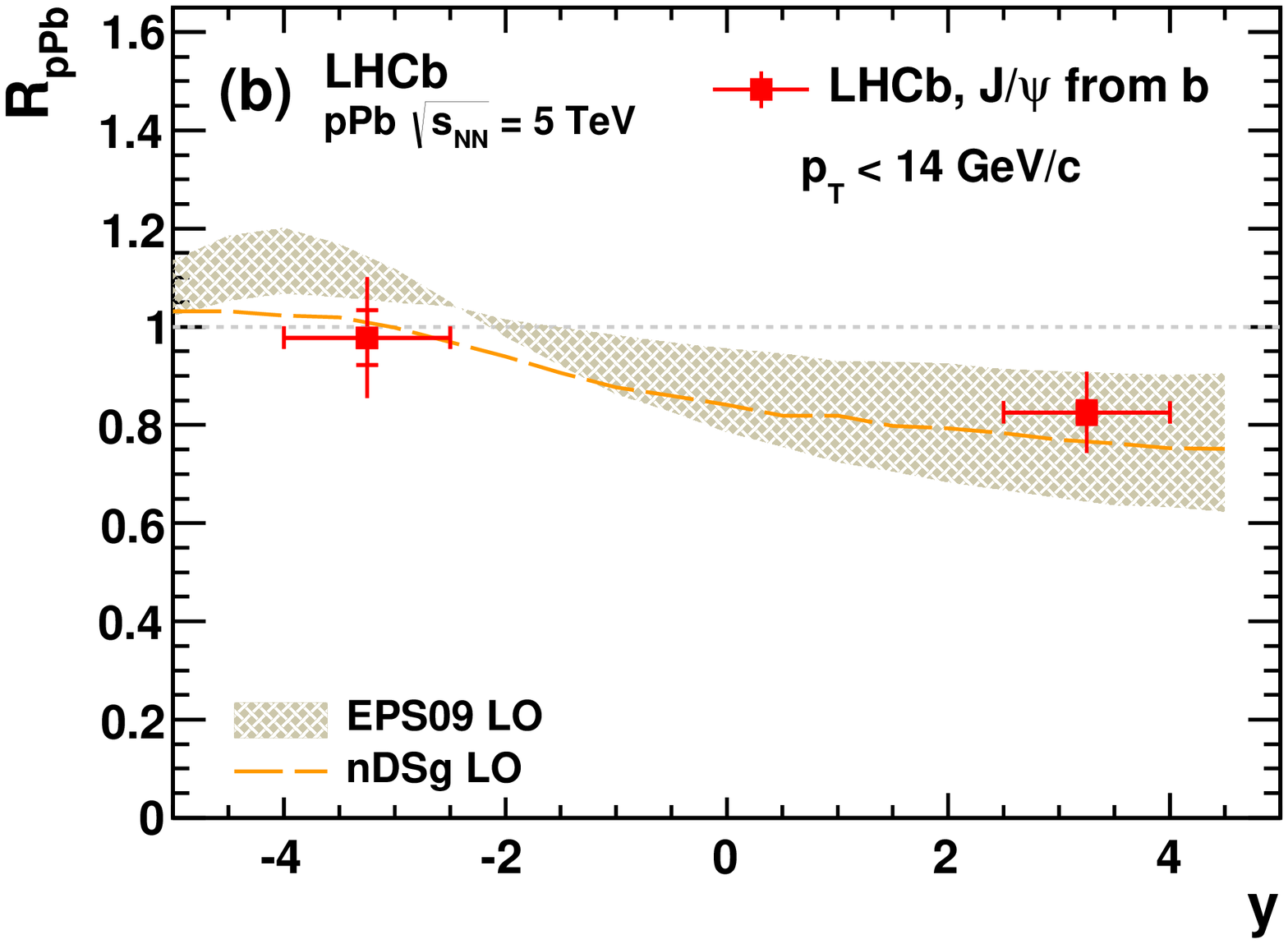}
\caption{Nuclear modification factor (top) and forward-backward production ratio (bottom) of (left) prompt $\jpsi$ and (right) $\jpsi$ from $b$ decay as a function of rapidity.}
\label{fig:Jpsi-RpA}
\end{figure}

\begin{figure}[htb]
\centering
\includegraphics[width=0.4\textwidth]{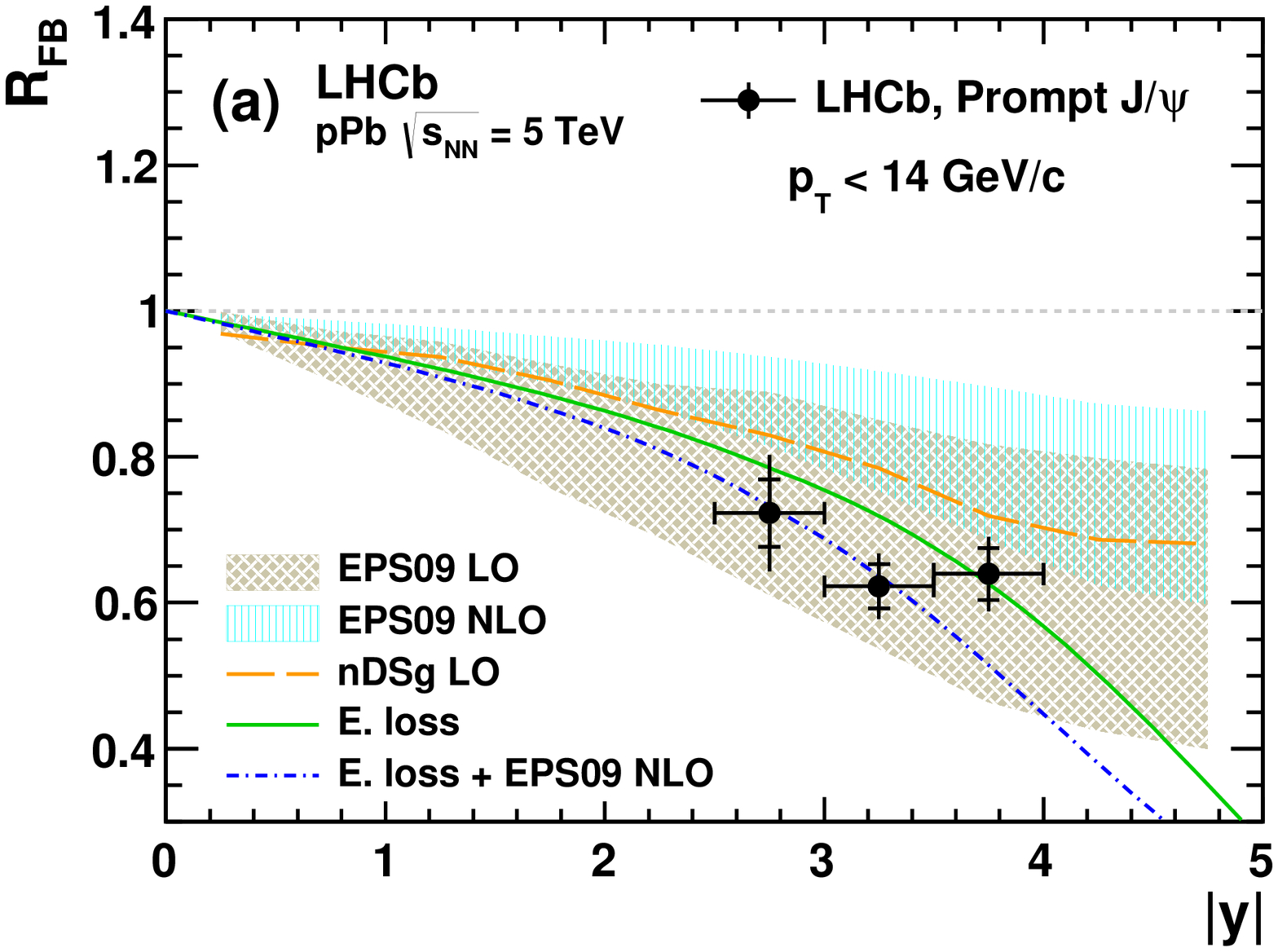}
\includegraphics[width=0.4\textwidth]{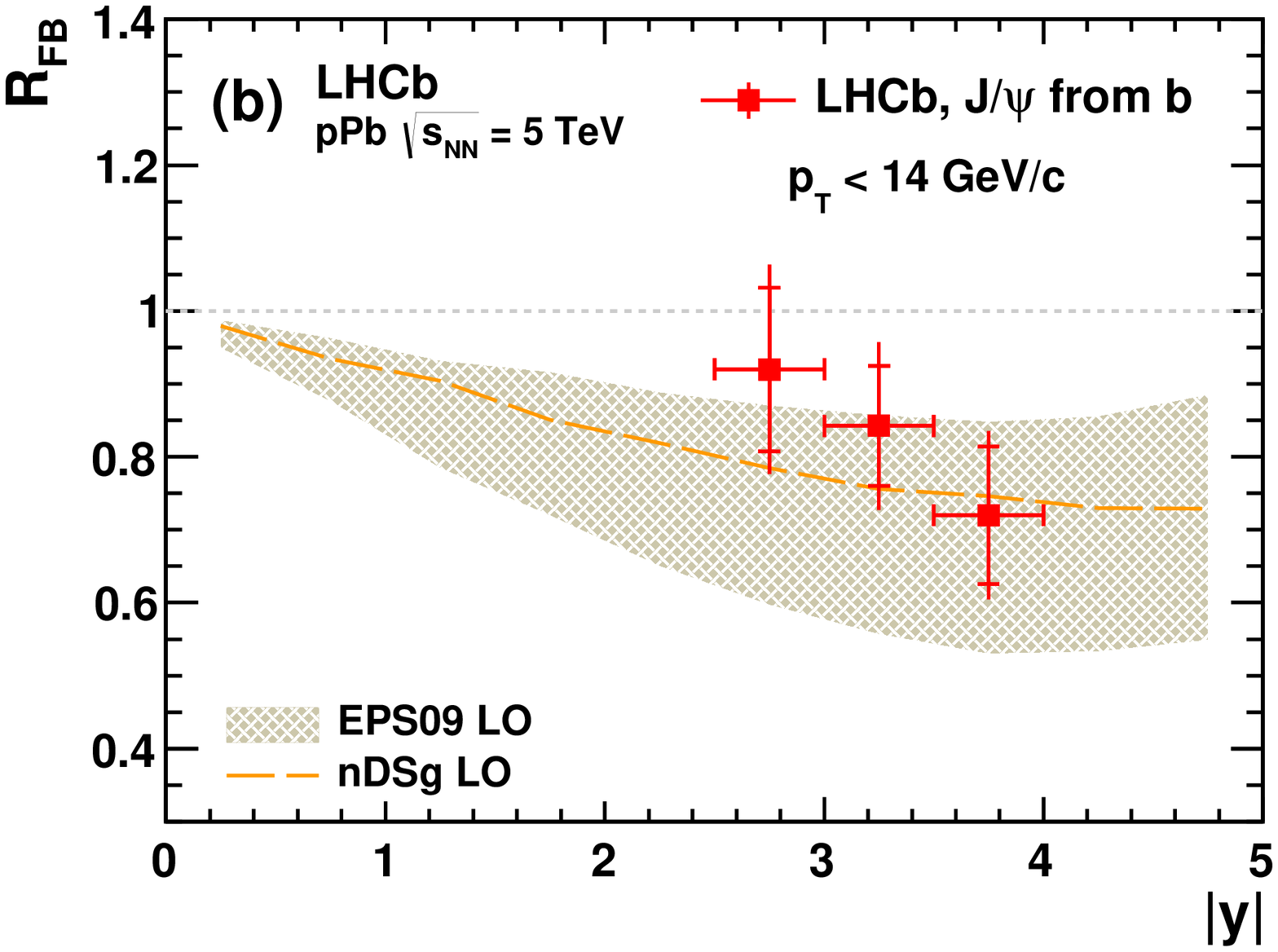}
\caption{Forward-backward production ratio of (left) prompt $\jpsi$ and (right) $\jpsi$ from $b$ decay as a function of rapidity.}
\label{fig:Jpsi-RFB}
\end{figure}

The measured nuclear modification factors for $\jpsi$ are shown in 
Figure \ref{fig:Jpsi-RpA}.
The $R_{\plead}$ are measured for prompt $\jpsi$ and $\jpsi$ from $b$ decays 
separately, because the latter reflects the CNM effects
on $b$ hadrons rather than $\jpsi$ itself. 
Strong suppression for prompt $\jpsi$ production in forward region is observed,
while the suppression of $\jpsi$ from $b$ is modest, indicating
the CNM effects on $b$ hadrons are less prominent.
The forward-backward production ratios
as a function of rapidity are shown in Figure \ref{fig:Jpsi-RFB}.
It confirms the suppression of prompt $\jpsi$ in the forward region, especially
at large rapidity; while the asymmetry of $\jpsi$ from $b$ decay is smaller.
The theoretical predictions~\cite{Ferreiro:2013pua,Albacete:2013ei,Arleo:2012rs,delValle:2014wha}  
generally agree with measured results, within 
large uncertainties.

\begin{figure}[htb]
\centering
\includegraphics[width=0.27\textwidth]{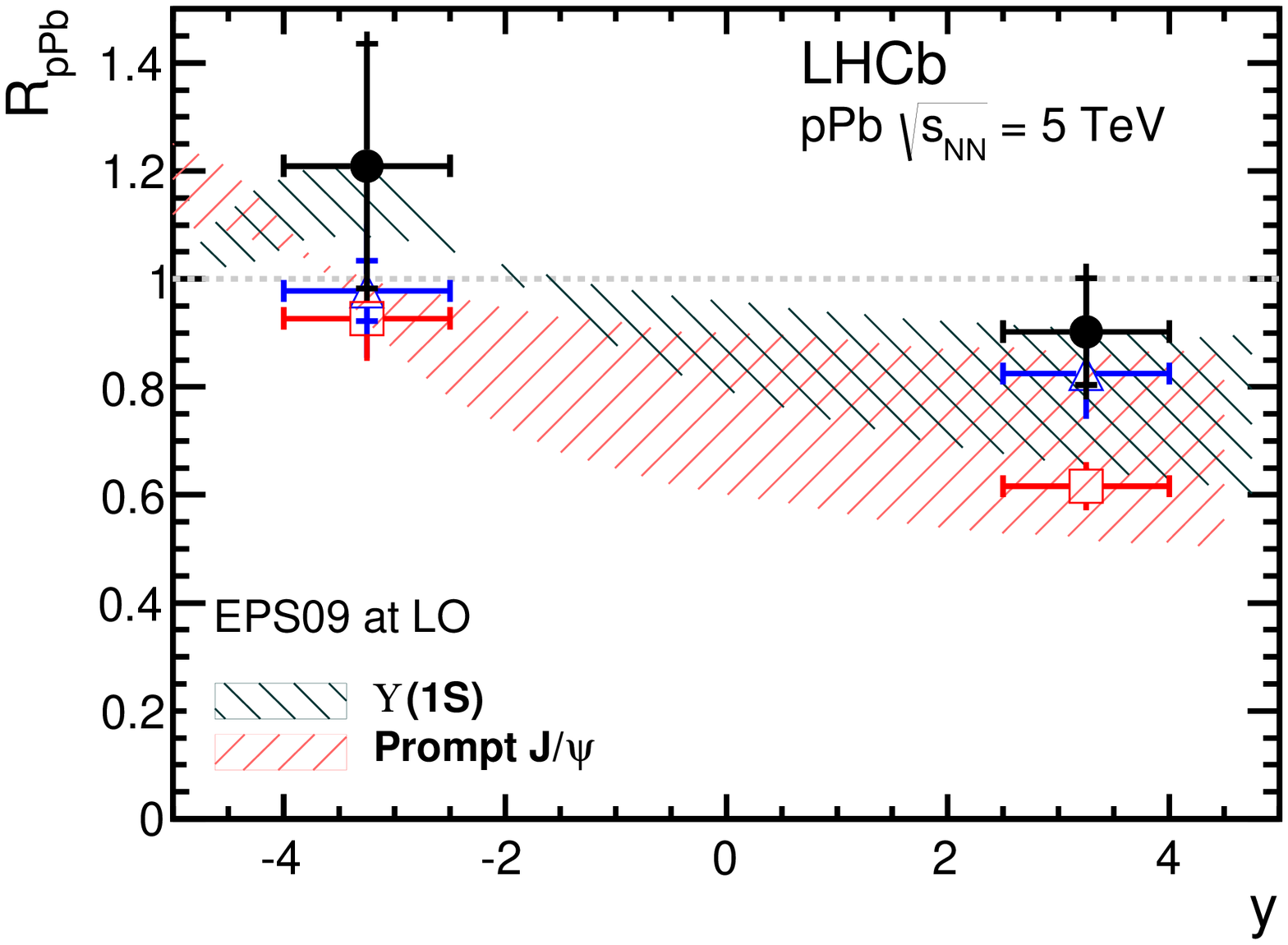}
\includegraphics[width=0.235\textwidth]{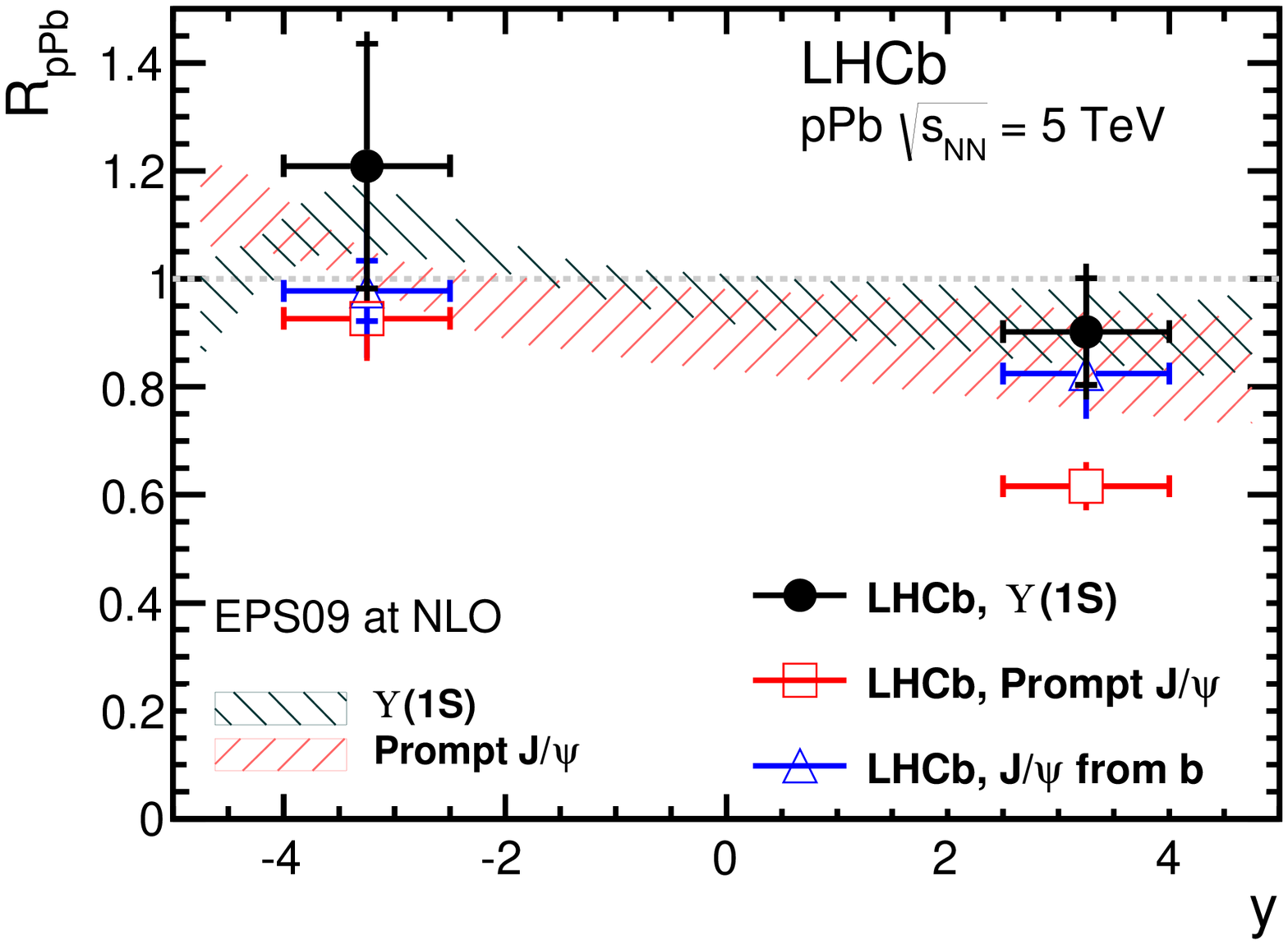}
\includegraphics[width=0.235\textwidth]{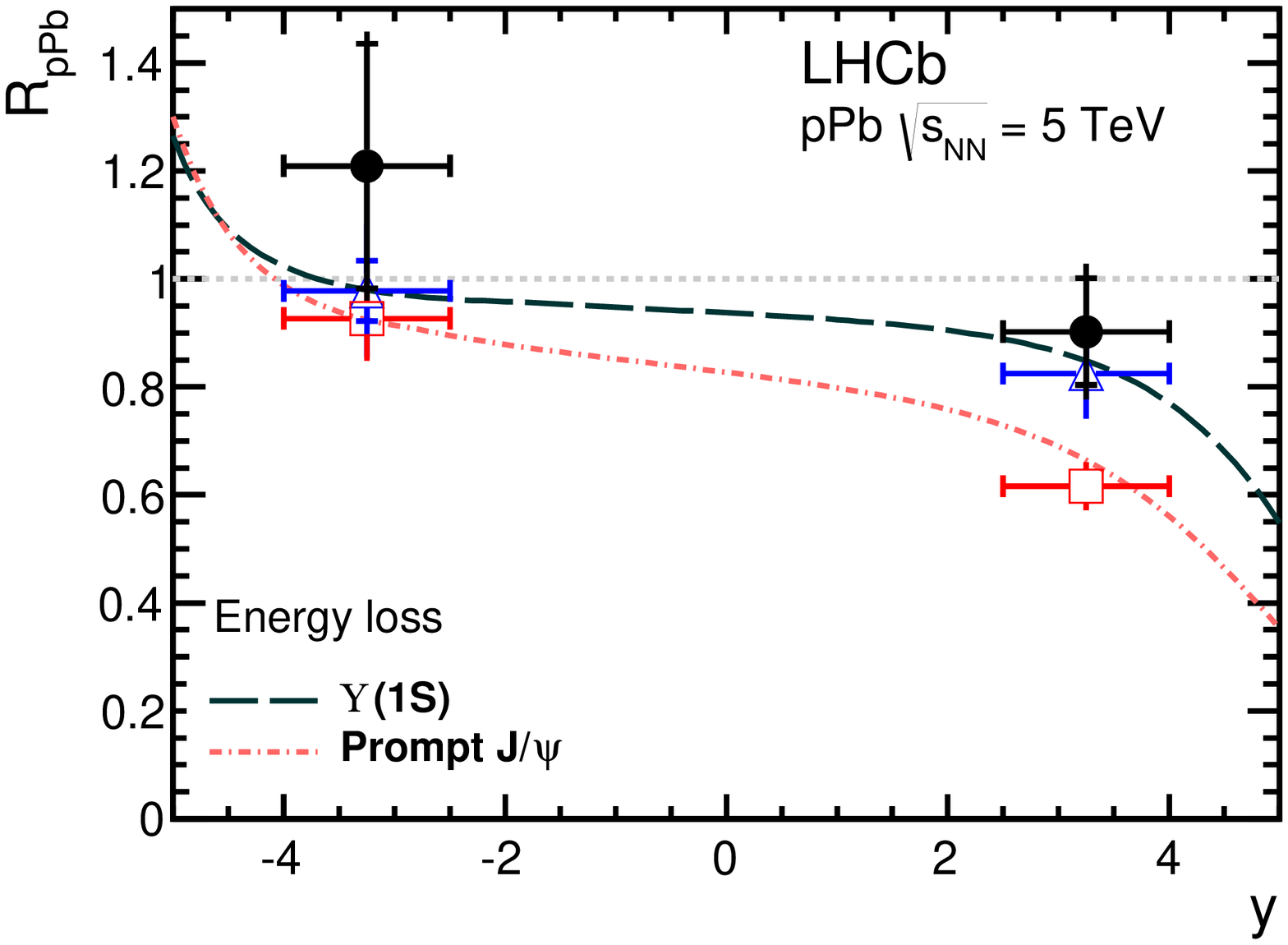}
\includegraphics[width=0.235\textwidth]{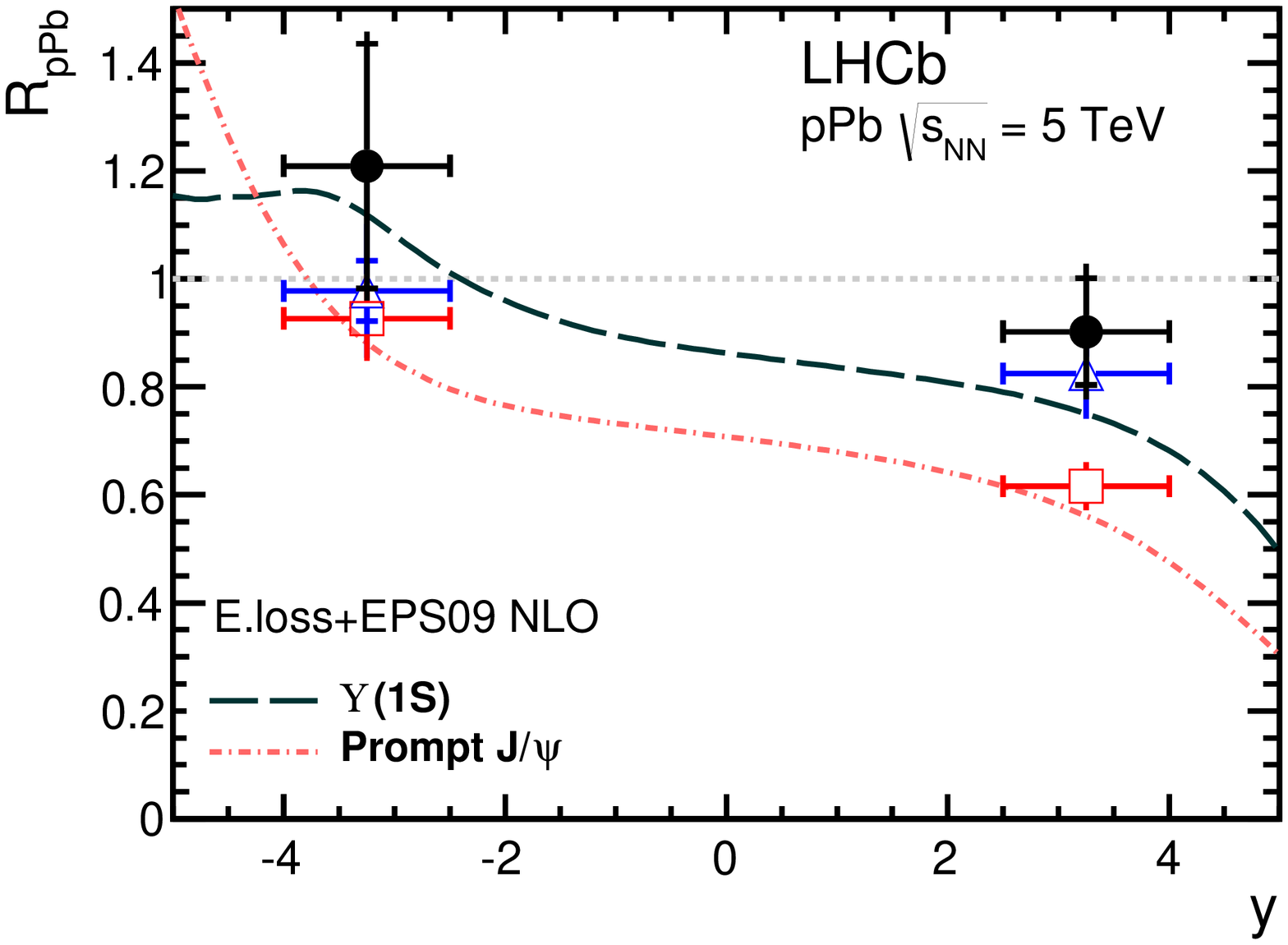}
\caption{Nuclear modification factors for prompt $\YoneS$ and $\jpsi$ as a function of rapidity, comparing with different theoretical calculations.}
\label{fig:Upsilon-RpA}
\end{figure}

\begin{figure}[htb]
\centering
\includegraphics[width=0.27\textwidth]{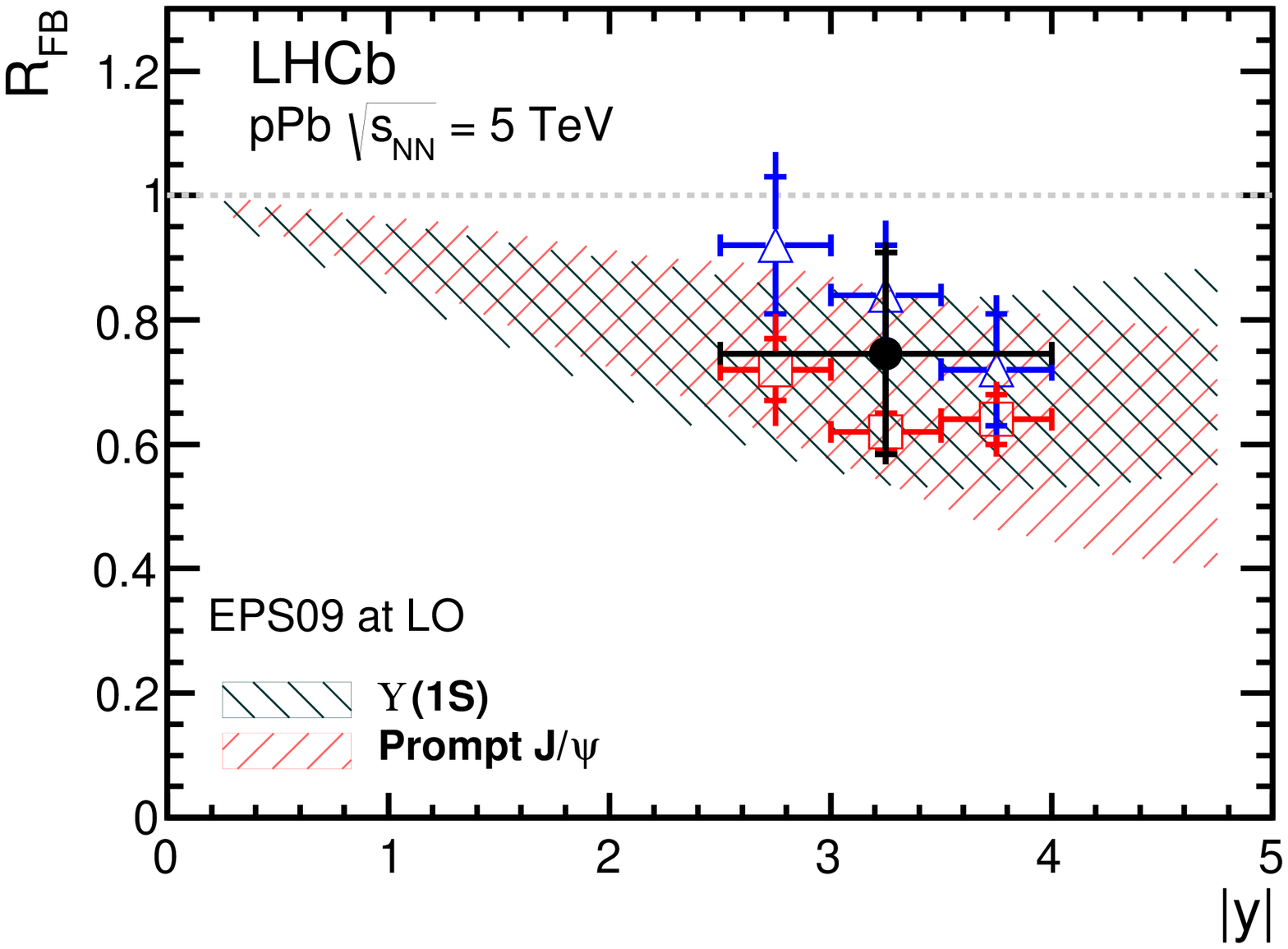}
\includegraphics[width=0.235\textwidth]{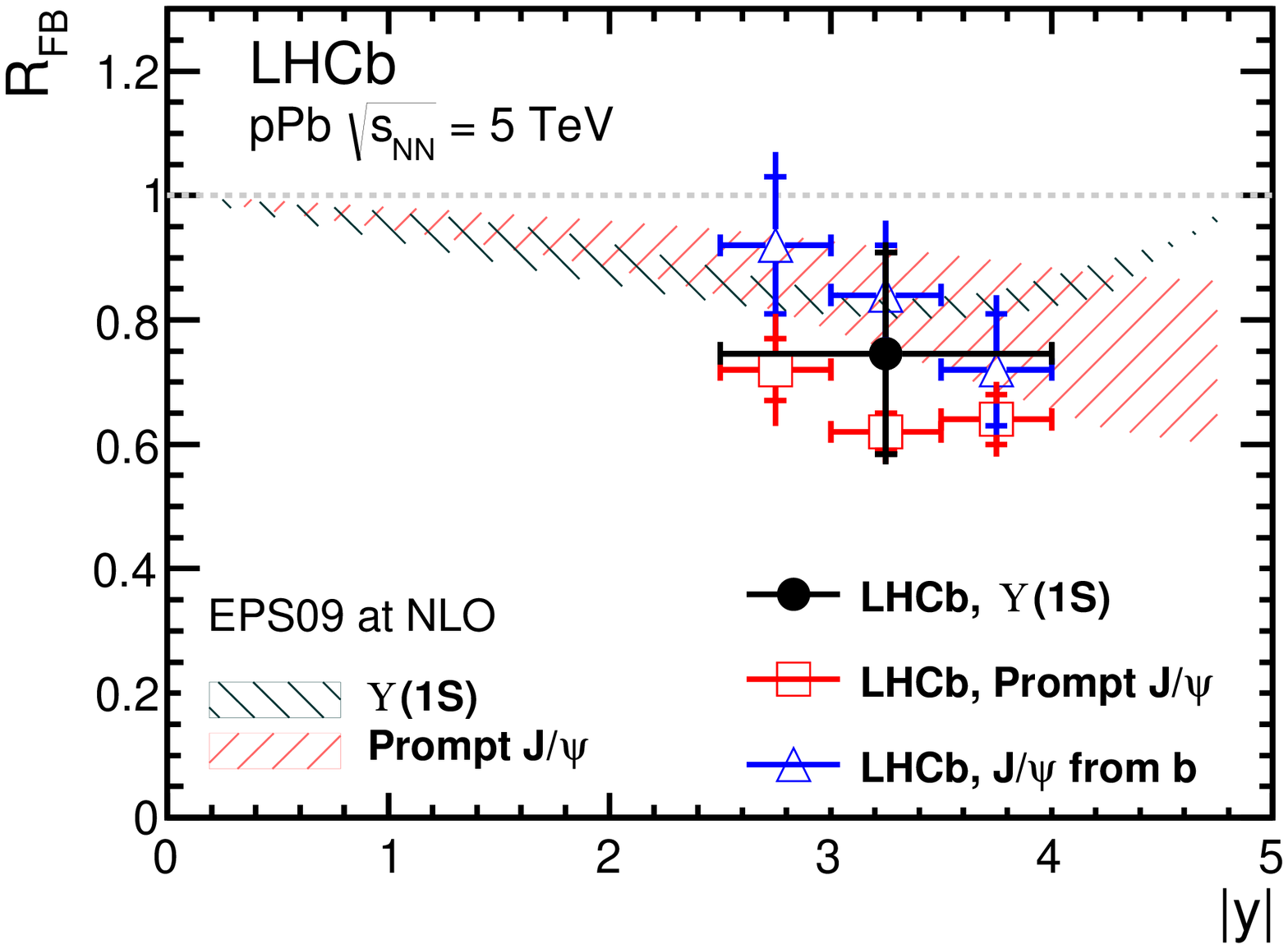}
\includegraphics[width=0.235\textwidth]{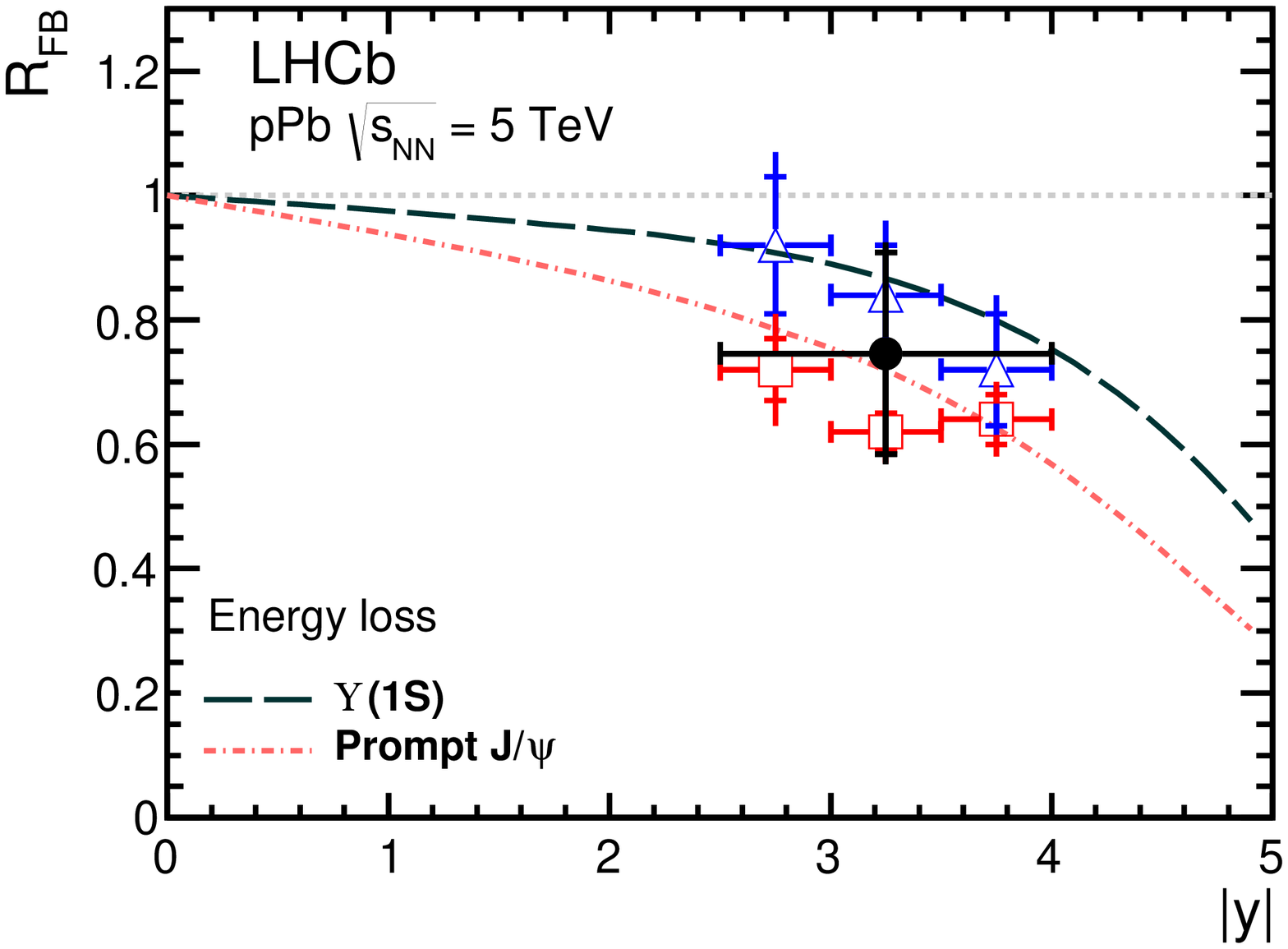}
\includegraphics[width=0.235\textwidth]{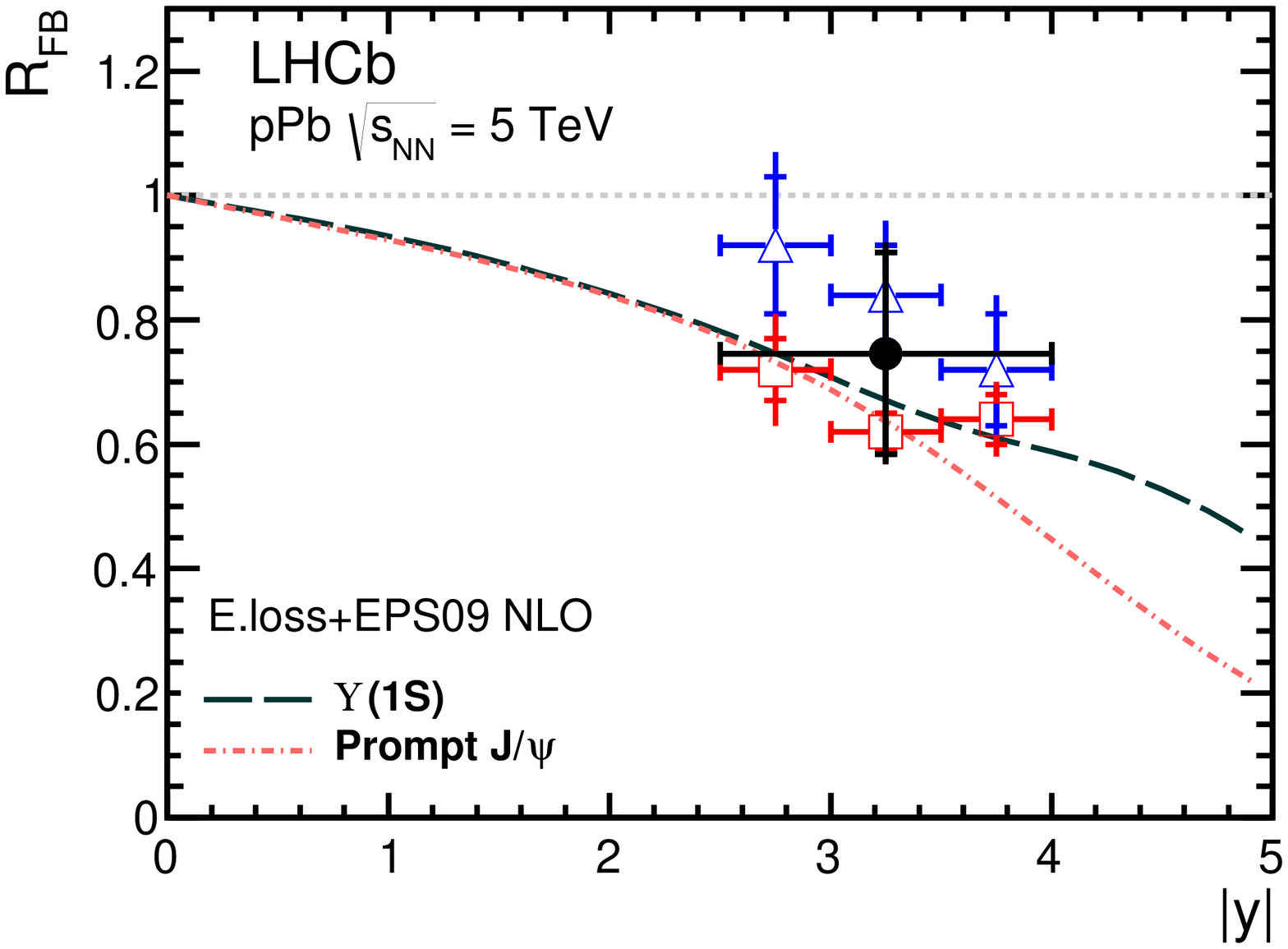}
\caption{Forward-backward production ratio for prompt $\YoneS$ and $\jpsi$ as a function of rapidity, comparing with different theoretical calculations.}
\label{fig:Upsilon-RFB}
\end{figure}

The measured $R_{\plead}$ and $R_{\mathrm{FB}}$ for $\YoneS$ as functions of
rapidity are shown in Figure \ref{fig:Upsilon-RpA} and \ref{fig:Upsilon-RFB} 
respectively, with $\jpsi$ results and each panel a theoretical calculation
~\cite{Albacete:2013ei,Arleo:2012rs,Ferreiro:2011xy} 
for comparison.
The suppression of $\YoneS$ production in the forward region is smaller than
prompt $\jpsi$, and close to $\jpsi$ from $b$ decay. This indicates that 
the CNM effects on $\YoneS$ is similar to that on $b$ hadrons. 
Due to large uncertainties all theoretical models are consistent with data.

\section{Conclusions}
The production of promt $\jpsi$, $\jpsi$ from $b$ decay and $\Upsilon$ mesons
at $\sqrt{s_{NN}} = 5 \tev$ in $\plead$ collisions are measured at LHCb, 
in rapidity range of $-5.0 < y < -2.5$ and $1.5 < y < 4.0$.
The cold nuclear matter effects are studied by measuring nulcear modification
factors $R_{\plead}$ and forward-backward production ratio for these mesons.
It is observed that prompt $\jpsi$ production is strongly suppressed 
in the forward region. The cold nuclear matter effects on $\YoneS$
and $\jpsi$ from $b$ hadrons are similar, both smaller than promt $\jpsi$.




\end{document}